\documentclass[doublecol]{epl2} 

\usepackage{amssymb}

\title{Feedback-controlled transport in an interacting colloidal system}
\shorttitle{Feedback-controlled transport} 

\author{Ken~Lichtner\inst{1} \and Sabine~H.~L.~Klapp\inst{1}}
\shortauthor{K.~Lichtner \etal}

\institute{                    
  \inst{1} Institute of Theoretical Physics, Secr. EW~7-1, Technical University Berlin, Hardenbergstr. 36, D-10623 Berlin, Germany.
}
\pacs{05.40.Jc}{Brownian motion}
\pacs{82.70.Dd}{Colloids}
\pacs{05.60.Cd}{Classical transport}

\abstract{
Based on dynamical density functional theory (DDFT) we consider a non-equilibrium system of interacting colloidal particles driven by a constant tilting force
through a periodic, symmetric ''washboard'' potential. We demonstrate that, despite of pronounced spatio-temporal correlations, the particle
current can be reversed 
by adding suitable feedback control terms to the DDFT equation of motion. We explore two distinct control protocols with time delay, focussing on either the particle positions or the
density profile. Our study shows that the DDFT is an appropriate framework to implement
time-delayed feedback control strategies widely used in other fields of nonlinear physics.}

\begin{document}

\maketitle

\section{Introduction}
Transport phenomena of Brownian particles in complex geometries are a topic receiving intense and continuos attention since decades \cite{reimann02,marche09,risken}.
A large number of studies has been devoted to transport in structured 1D systems such as colloids or biomolecules in microchannels \cite{ros05}, colloids 
in optical potentials \cite{lee06,blickle07}, or cold atoms in optical lattices \cite{gommers08}.
Theoretical studies of such systems have predicted spectacular 
effects such as ratchet mechanisms in systems with asymmetric spatial potential
 \cite{reimann02}, giant diffusion \cite{costantini99,reimann01,lee06,blickle07} and dispersionless transport \cite{lindenberg07} in symmetric systems under constant 
 external bias ("tilted washboards"), and the negative mobility effect \cite{ros05,eichhorn10}. Many of these effects have also been observed 
experimentally (see, e.g., \cite{lee06,blickle07,eichhorn10,evsti08}), often
involving colloidal systems. A related topic is how these non-equilibrium phenomena can be manipulated by control forces
\cite{cao04}. Particularly promising are {\em feedback control} schemes which depend on the state
of the system. A special case is the time-delayed feedback control method suggested by Pyragas \cite{pyragas92}, 
where the control term involves the difference between an output variable (the control target) at time $t$ and its value at time $t-\tau$, with $\tau$ being the delay time. This 
method is particularly suitable to stabilize certain, otherwise unstable (periodic) states. Moreover, a time delay naturally occurs in experiments involving feedback control
due to the lag between the collection of information and the feedback. Indeed, time-delayed feedback control is nowadays used in a broad variety of 
non-linear systems such as lasers, neural dynamics, and excitable macroscopic
media \cite{Schoellbuch}. In the area of 1D transport, the method has already been applied, on a theoretical level, to Brownian motors (driven by an unbiased, time-periodic force) \cite{wu06}, systems rectified by delayed correlated noise \cite{borromeo06}, and flashing ratchets (involving asymmetric, time-dependent potentials)\cite{cao04,craig08}.  A first experimental realization of a feedback-controlled flashing ratchet already exists  \cite{lopez08}. 
Very recently, feedback strategies with delay have also been explored, on the basis of Langevin equations  \cite{hennig1,hennig2}, as a tool to manipulate the particle current in tilted washboard potentials.
However, despite intense research we are still far away from a full understanding of the usefulness of control schemes in transport processes 
and its potential applications in biology and nanotechnology. 

One open, yet very important point is the role of particle interactions, investigations of which have only started recently (see, e.g., \cite{evsti09}).
In the present letter we show, for the first time, that control in an interacting, driven, overdamped system 
can be well implemented within the framework of dynamical density functional theory (DDFT) \cite{marconi1,marconi2,archer04}.
In the DDFT the basic dynamic variable is the time-dependent, continuous density field, 
where the microscopic interactions determining the underlying system of discrete Brownian particles
enter via a free energy functional. Therefore, DDFT has a bridging position between microscopic (Langevin-equation based) models, on one hand, 
and mesoscopic, hydrodynamic models for transport of continuous phases (see, e.g., \cite{john07}), on the other hand; a connection, which is also highlighted by a recent derivation of the DDFT via projector operator techniques \cite{espanol09}. 
In the last years, DDFT has been applied to a variety of driven systems such as
colloids in unstable traps \cite{rex08} and sedimenting colloids \cite{royall07}. Moreover, in a very recent work \cite{pototsky10}, DDFT is employed to study attracting colloidal particles in 1D (time-dependent) ratchet potentials. Here we apply the method to a system of soft, repulsive colloids in a static, tilted washboard potential. We focus
on the impact of time-delayed feedback control. Indeed, we find
that the current in our strongly correlated system can be efficiently controlled via time-delayed feedback 
schemes focussing either on the average particle position or directly on the density profile.
\section{Model and Results}
Our model system consists of $N$ colloidal particles in an one-dimensional channel of length $L$. Particles at positions $z_1$ and $z_2$ in the channel interact 
via a repulsive Gaussian potential $V^{\mathrm{GCM}}(z_1-z_2)=\varepsilon_0\exp[-(z_1-z_2)^2/\sigma^2]$
(with $\varepsilon_0>0$), a typical coarse-grained potential modeling a wide class of soft, partially penetrable macroparticles (e.g., polymer coils)
with effective (gyration) radius $\sigma$ \cite{likos01,louis00}. The particles are subject 
to a spatially periodic, symmetric (''washboard'') potential $U^{\mathrm{wb}}(z)=U_0\sin^2(k z)$, 
where $k$ defines the wavelength and $U_0$ is the amplitude. 

\begin{figure}
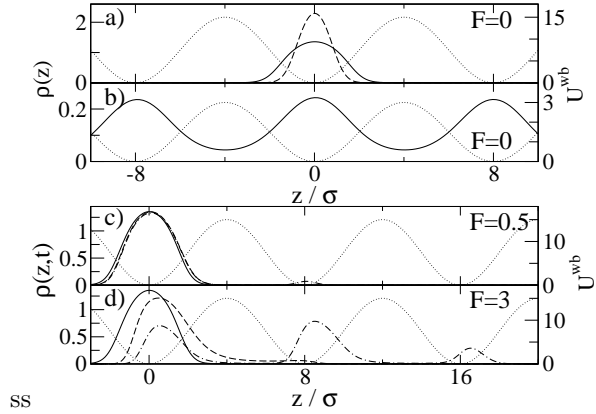

\onefigure[width=7.5cms]{fig1.eps}
\caption{Density profiles as a function of the particle position at $\varepsilon=4$ and $N=4$. The washboard potential $U^{\mathrm{wb}}(z)$ (with $k=\pi/(8\sigma)$)
is indicated by the dotted lines. Top: profiles for non-driven systems ($F=0$) at a) $U=15$ and b) $U=3$. In a) we have included 
a profile for $\varepsilon=1$ (dashed line). Bottom: profiles for driven systems at c) $F=0.5$ and d) $F=3$ and times $t=0$ (solid), $t=\tau_{\mathrm{B}}$ (dashed) 
and $t=10\tau_{\mathrm{B}}$ (dot-dashed). The total length of the channel is $L= 60\sigma$.}
\label{density}
\end{figure}

Typical distributions of the system in the absence of an external drive 
are illustrated in fig.~\ref{density}a and fig.~\ref{density}b. Specifically, we consider the space-dependent 
one-particle density $\rho(z)=\langle \sum_{i=1}^N\delta\left(z-z_i\right)\rangle$
(with $\langle\ldots\rangle$ being an ensemble average) at
two values of the interaction strength $\varepsilon=\varepsilon_0/k_{\mathrm{B}}T$, where 
$k_{\mathrm{B}}$ and $T$ are Boltzmann's constant and temperature, respectively. The particle number is fixed
to $N=\int_{-L/2}^{L/2}dz \rho(z,t)=4$.
The data have been obtained numerically using the DDFT formalism described below, 
with the initial configuration ($t=0$) being a single density peak centered in one potential well. The data in fig.~\ref{density}a
pertain to a washboard potential of strength $U=U_0/k_{\mathrm{B}}T=15$. Here, thermal fluctuations are too weak to allow for 
barrier crossing within the observation time and, consequently, the particles remain localized in the potential well.
This is true both for strong interactions ($\varepsilon=4$) and in a
small-coupling case ($\varepsilon=1$), where the narrower peak of $\rho(z)$ suggests an even more pronounced localization. Figure~\ref{density}b shows, for comparison, a profile corresponding to $U=3$ (and $\varepsilon=4$, $F=0$), yielding a fluid-like situation
with the density becoming non-zero everywhere. Within our calculations, we did not identify a threshold value of $U$ separating the two regimes, consistent with
the continuous dependence of the equilibrium mobility on $U$ in the corresponding (exactly solvable) single-particle problem \cite{risken}.

From now on we focus on the more interesting localized situation depicted in fig.~\ref{density}a. Here, an effective motion of the particles can be induced by a constant 
tilting force ${\mathbf F}^{\mathrm{bias}}=F_0 \hat{\mathbf{z}}$ (with $\hat{\mathbf{z}}$ being the unit vector in $z$-direction), corresponding to a linear potential $U^{\mathrm{bias}}(z)=-F_0z$. We choose $F_0>0$ such that the
particles move preferentially to the right. The non-equilibrium dynamics of 
the driven system is investigated via dynamical density functional theory (DDFT) \cite{marconi1,marconi2,archer04,espanol09}, where the central quantity is the time-dependent, one-particle density of the particles, $\rho(z,t)$. By construction, DDFT assumes the dynamics to be overdamped, i.e., inertial effects are neglected. 
The exact Smoluchowski equation for $\rho(z,t)$ is then approximated such that non-equilibrium two-particle correlations at time $t$ are set to those
of an equilibrium system with density $\rho(z,t)$. Neglecting, moreover, hydrodynamic interactions, one obtains the key DDFT equation 
\cite{marconi1,marconi2,archer04,espanol09}
\begin{equation}
\label{ddft}
 \Gamma^{-1}\frac{\partial \rho(z,t)}{\partial t}=\nabla \left[\rho(z,t)\nabla \frac{\delta {\cal F}[\rho(z,t)]}{\delta \rho(z,t)}\right],
\end{equation}
where $\Gamma$ is a mobility coefficient (i.e., $\Gamma=D_0/k_{\mathrm{B}}T$ with $D_0$ being the short-term diffusion coefficient), and ${\cal F}[\rho]$ is a free energy density functional determining the effective ''current'' 
$\mathbf{j}=-\Gamma\rho\nabla\left(\delta {\cal F}[\rho]/\delta \rho\right)$. Specifically, ${\cal F}={\cal F}^{\mathrm{id}}+{\cal F}^{\mathrm{int}}+{\cal F}^{\mathrm{ext}}$ where
${\cal F}^{\mathrm{id}}[\rho]=k_{\mathrm{B}}T\int\upd z \rho(z,t)\left[\ln\left(\Lambda^3\rho(z,t )\right)-1\right]$
is the ideal part (with $\Lambda$ being the 
thermal wavelength),
${\cal F}^{\mathrm{ext}}=\int\upd z \rho(z,t)\left(U^{\mathrm{wb}}(z)+U^{\mathrm{bias}}(z)\right)$ is the external field contribution,
and ${\cal F}^{\mathrm{int}}$ accounts for the colloidal interactions. Here we employ the mean-field (MF) approximation, that is,
${\cal F}^{\mathrm{int}}=(1/2)\int\upd z_1\int\upd z_2\rho(z_1,t)V^{\mathrm{GCM}}(|z_1-z_2|)\rho(z_2,t)$. Due to the penetrable nature
of the Gaussian potential (which allows an, in principle, infinite number of neighbors) the MF approximation
is known to become quasi-exact in the high-density limit and yields reliable results
even at low and moderate densities \cite{louis00}. 

The impact of the external drive on the density profile is illustrated in fig.~\ref{density}c and fig.~\ref{density}d, where we consider two values of the 
driving strength $F=F_0\sigma/k_{\mathrm{B}}T$. The numerical calculations have been performed in channels with closed boundaries, such that
any drift motion is, strictly speaking, of transient character. However, by choosing channels of sufficient lengths ($L\geq 60\sigma$) we have 
ensured that all shown profiles are free of boundary effects.

\begin{figure}
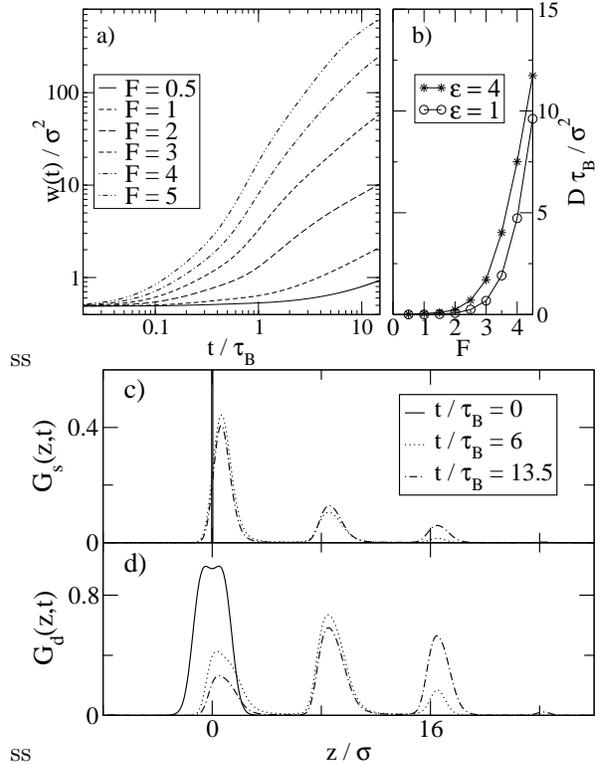

\onefigure[width=7.5cms]{fig2ab.eps}
\onefigure[width=7.5cms]{fig2cd.eps}
\caption{a) Mean-square displacement as function of time at several driving strengths ($\varepsilon=4$, $N=4$). b) Corresponding diffusion constants. 
Included are date for $\varepsilon=1$. Parts c) and d) show the self and distinct part of the van Hove function
at $F=3$, $\varepsilon=4$, and some characteristic times.}
\label{vanhove}
\end{figure}

In the small-drive situation depicted in fig.~\ref{density}c ($F=0.5$), the peak barely moves within the time range considered, indicating
that the probability for the particles to jump over the barriers is still small. This changes
at $F=3$ (fig.~\ref{density}d) where the drive causes a shift of the entire density peak to the right (here and in the following, time is measured in units of the Brownian timescale $\tau_{\mathrm{B}}=\sigma^2/(\Gamma k_{\mathrm{B}}T)$, which is of the order of
$10^{-9}$s for typical colloids). At the same time, the distribution
broadens in the sense that now several potential wells are (potentially) occupied by particles.
The influence of $F$ is also reflected by the time-dependence of 
the mean-squared displacements (MSD), $w(t)=N^{-1}\sum_{i=1}^N\langle \left(z_i(t)-z_i(0)\right)^2\rangle$. Since, within the DDFT formalism, we do not have direct access
to the particle positions, we obtain the MSD rather via 
the relation $w(t)=\int_{-\infty}^{+\infty} \upd z z^2 G_{\mathrm{s}}(z,t)$. Here, $G_{\mathrm{s}}(z,t)$
is the self-part of the van Hove correlation function measuring the probability that a particle moves over a distance $z$ during time $t$.
We calculate this function, as well as its distinct counterpart $G_{\mathrm{d}}(z,t)$ measuring
the time dependence of two-particle correlations, within the DDFT formalism via the test particle method \cite{archer07}. Some numerical results for the resulting MSD are given in 
fig.~\ref{vanhove}a; corresponding diffusion constants $D=\lim_{t\to\infty}w(t)/2t$ are plotted in fig.~\ref{vanhove}b.
Within the time range considered, we find diminutive values of $D$ 
for driving strength $F\lessapprox 0.5$, consistent with the behavior of the density profile
in fig.~\ref{density}c. Upon further increase of $F$,
the MSD at intermediate times first displays sub-diffusion (where $w(t)$ increases slower than linearly with $t$) and then
superdiffusion characterized by much faster growth.
In the long-time limit, the systems reach true diffusive behavior with $w(t)\propto t$ as expected
from the Brownian friction incorporated in the DDFT. 
Exemplary data for the functions $G_{\mathrm{s}}(z,t)$ [at $F=3$] are plotted in fig.~\ref{vanhove}c. The appearance of several peaks
in $G_{\mathrm{s}}(z,t)$ reflects the existence of several ''populations'' of particles, one consisting of particles that remain in their original potential valley and the other
ones consisting of moving particles. Clearly, the self-van Hove function strongly deviates from the gaussian behavior expected in a ''normal'', liquid-like 
diffusing system.
Coming back to fig.~\ref{vanhove}b we note  
that the maximum value of $F$ considered here is still smaller
than the "critical'' force $F_{\mathrm{crit}}=\mathrm{max}(d U^{\mathrm{wb}}/dz)\approx 5.9$, 
beyond which the potential barriers are eliminated, and a single particle can slide freely.
Therefore, we do not see the maximum and subsequent decrease of $D$ expected in the vicinity of $F_{\mathrm{crit}}$ \cite{costantini99,reimann01} 
(indeed, investigation of that range is hindered
by the finite length of our system which eventually yields boundary effects). We recall in this context that the particles in our system are {\em interacting}
such that one may expect deviations of the behavior of $D$  from the uncorrelated case \cite{evsti09}. Indeed, as indicated in fig.~\ref{vanhove}b, a decrease of the coupling
strength yields a shift of the curve $D(F)$ towards larger driving strength. 
Thus, the repulsive interactions in our model support the particles in crossing the barriers.
Finally, fig.~\ref{vanhove}d shows the distinct van Hove function, $G_{\mathrm{d}}(z,t)$, as calculated by the
test particle method \cite{archer07}. At $t=0$, where $G_{\mathrm{d}}$ is proportional to the usual pair distribution function, the correlations are restricted to the first potential well in which the particle
where confined initially (note the correlation hole at $t=0$). At later times $G_{\mathrm{d}}(z,t)$ develops additional peaks in the neighboring
potential valleys at $z>0$, indicating pronounced spatio-temporal correlations in the driven system.

We now aim at manipulating the dynamics by time-delayed feedback control, which involves the difference between an appropriate
system variable (the control target) at time $t$ and its value at time $t-\tau$. The use of 
such closed-loop strategies in 1D systems subject to tilted washboard potentials has been previously explored, e.g., in Refs.~\cite{hennig1,hennig2}. These studies
investigated {\em non-interacting} Brownian particles by direct numerical solution of the corresponding Langevin equations.
Here we consider an interacting colloidal system and the dynamics is described by the DDFT equation~(\ref{ddft}).

We explore two distinct control strategies. Within the first protocol, 
which is similar in spirit to that in \cite{hennig1}, the control target is the first moment of the density corresponding to the average particle position
$\bar{z}=\int_{-\infty}^{+\infty} \upd z z \rho(z,t)$. 
We note that particle positions and thus, their average, are measurable variables
in experiments of colloidal transport (see, e.g., refs.~\cite{lee06,blickle07}). To implement the feedback control
we supplement the external potential entering in the free energy functional ${\cal F}^{\mathrm{ext}}$
(see text below eq.~(\ref{ddft})) by a linear control potential $U^{\mathrm{c}}_1(z,t;\tau)=-zF^{\mathrm{c}}_1(t;\tau)$ where
\begin{equation}
\label{f_control1}
F^{\mathrm{c}}_1(t,\tau)=-K_0\left (1-\tanh\left[\bar{z}(t)-\bar{z}(t-\tau)\right]\right)
\end{equation}
is a spatially homogeneous control force. In the DDFT equation~(\ref{ddft}) this force yields a current 
${\mathbf{j}}^{\mathrm{c}}=-\Gamma\rho(z,t)F^{\mathrm{c}}(t,\tau)$. Note the restriction
$-2K_0\leq F^{\mathrm{c}}_1\leq 0$, where $K_0$ is assumed to be positive. In the numerical calculations we switch on the control force at
$t_{\mathrm{start}}=1.5\tau_{\mathrm{B}}$ (i.e., $F^{\mathrm{c}}_1(t<t_{\mathrm{start}})=0$).

\begin{figure}
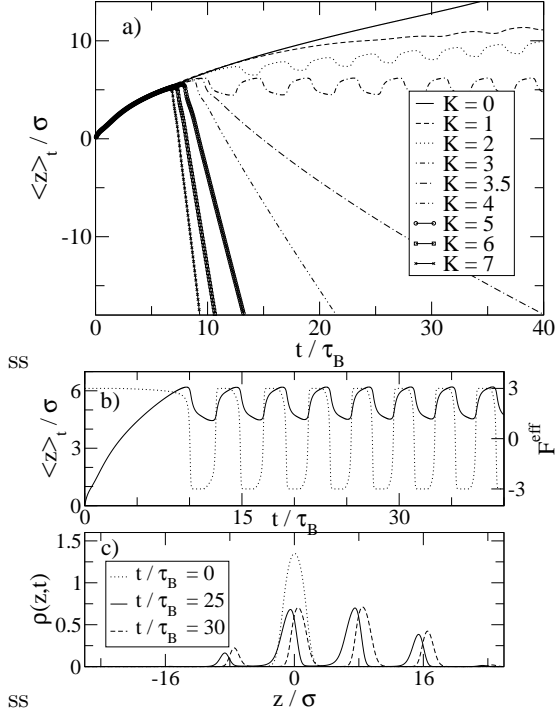

\onefigure[width=7cms]{fig3a.eps}
\onefigure[width=7cms]{fig3bc.eps}
\caption{a) Normalized first moment as a function of time for various control amplitudes $K$ (see eq.~(\ref{f_control1})). b)
First moment (solid line) and effective bias (dashed) at $K=3$. c) Density profiles for various times ($F=3$, $\tau=\tau_{\mathrm{B}}$, $t_{\mathrm{start}}=1.5\tau_{\mathrm{B}}$).}
\label{control1}
\end{figure}

The impact of the control force $F^{\mathrm{c}}_1$ on the normalized first moment $\langle z\rangle_t=\bar{z}(t)/N$ is illustrated in fig.~\ref{control1}a, 
where we have chosen $F=3$ and $\tau=\tau_{\mathrm{B}}$. 
This delay time corresponds roughly to an {\em intrinsic} time scale of the system. Indeed,
as seen from fig.~\ref{vanhove}a, $t=\tau_{\mathrm{B}}$ is within the time range where the crossover from sub-diffusive into diffusive behavior
of the MSD of the uncontrolled system occurs.

In the absence of control ($K=K_0\sigma/k_{\mathrm{B}}T=0$) the first moment plotted in fig.~\ref{control1}a
just increases with $t$, reflecting
the rightward motion expected at $F=3$ (see fig.~\ref{density}d).
The slope of the function $\langle z\rangle_t$ at large $t$ may be interpreted as an average velocity 
$v=\lim_{t\to\infty}\left(d\langle z\rangle_t/dt\right)$.
Increasing $K$ from zero, the velocity first decreases until the peak motion stops (i.e., the time-average of
$\langle z\rangle_t$ becomes constant) at $K=3$. This value corresponds to a balance between control force and bias. Even larger control amplitudes
then result in a significant backward motion, i.e., $\langle z\rangle_t$ and $v$ become negative. 

\begin{figure}
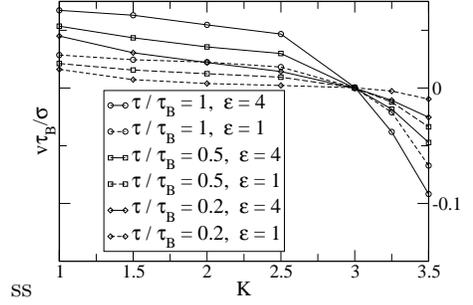

\onefigure[width=5.7cms]{fig4.eps}
\caption{Long-time velocity as function of $K$ for several values of $\tau$ and $\varepsilon$ ($F=3$, $t_{\mathrm{start}}=1.5\tau_{\mathrm{B}}$). Lines are guides for the eye.}
\label{control2}
\end{figure}

We now consider in more detail the time-dependence of the feedback control. 
First, a significant influence 
on $\langle z\rangle_t$ appears only at relatively large times $t\gg t_{\mathrm{start}}$ (which depend, in turn, on the actual value of $K$).
At earlier times,
the density peak moves so quickly that 
$\bar{z}(t)\gg\bar{z}(t-\tau)$ and the $\tanh$-function in eq.~(\ref{f_control1}) approaches $1$, yielding
$F^{\mathrm{c}}_1(t,\tau)\approx 0$. With the ''slow-down'' of $\langle z\rangle_t$
at somewhat later times (visible also at $K=0$), the argument of the $\tanh$ decreases. 
Thus, the control sets in, yielding an {\em effective} biasing force $F^{\mathrm{eff}}(t)=F^0+F^{\mathrm{c}}_1(t;\tau)\leq F_0$. 
The behavior of $F^{\mathrm{eff}}(t)$ and the function $\langle z\rangle_t$ is shown
in fig.~\ref{control1}b for the ''balanced'' case
$K=3$. One sees that the control becomes effective at $t\approx 10\tau_{\mathrm{B}}$. After that, the first moment displays an oscillating behavior changing
between small backward motion and forward motion. These oscillations yield, in turn, oscillations of $F^{\mathrm{eff}}\sigma/k_{\mathrm{B}}T$ 
between its maximum value, $F=3$, and its minimum $F-2K=-3$. 
As a consequence, the overall motion stops. Interestingly, the
oscillations (which seem to persist in the long-time limit) 
have a period of about $5\tau_{\mathrm{B}}$, that is, much larger than the delay time ($\tau=\tau_{\mathrm{B}}$). Density profiles
related to one ''cycle'' of $\langle z\rangle_t$ and $F^{\mathrm{eff}}(t)$ 
are plotted in fig.~\ref{control1}c, where $t/\tau_{\mathrm{B}}=25$ and $30$ correspond roughly to the minimum and maximum of $F^{\mathrm{eff}}$, respectively
(see fig.~\ref{control1}b). It is seen that the entire particle distribution
is shifted with the periodic changes of $F^{\mathrm{eff}}$. These shifts are accompanied by changes in the peak shapes, which become most asymmetric
when $F^{\mathrm{eff}}$ is minimal, i.e., the control is maximal ($t/\tau_{\mathrm{B}}=25$).
To complete the picture, we plot in fig.~\ref{control2} the long-time velocity $v$ (averaged over the oscillations of $\langle z\rangle_t$, if present)
as function of the
control amplitude. We have included data for different delay times $\tau$ and different interaction (i.e., repulsion) strengths $\varepsilon$.
All systems considered display a clear current reversal at $K=3$, where the velocity $v$ changes from positive to negative values irrespective of $\varepsilon$ and $\tau$.
These parameters, however, do have an
impact on the {\em magnitude} of the velocities in the two regimes $K<3$ and $K>3$. Specifically, reduction of $\varepsilon$ (at fixed $\tau$) yields a decrease of $v$ as compared
to the case $\varepsilon=4$. Similarly, $v$ decreases in magnitude when the delay time decreases (at fixed $\varepsilon$)
from $\tau=\tau_{\mathrm{B}}$ towards $\tau=0.2\tau_{\mathrm{B}}$. In other words, the time delay {\em supports} the current reversal in the parameter range considered.
We also note that all of these results are robust, on a qualitative level, against slight changes of the control
protocol, such as a reversal the argument of the transcendental function in eq.~(\ref{f_control1}).

\begin{figure}
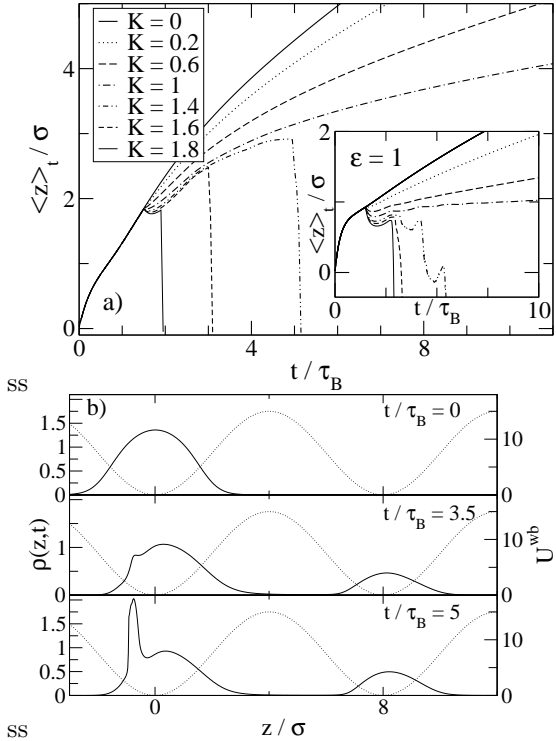

\onefigure[width=7cms]{fig5a.eps}
\onefigure[width=7cms]{fig5b.eps}
\caption{a) Normalized first moment as a function of time and various control amplitudes $K$ for the second
control loop (see eq.~(\ref{u_control3})) and $\varepsilon=4$. The inset in a) contains data for $\varepsilon=1$ at otherwise same parameters
$F=3$, $\tau=\tau_{\mathrm{B}}$, and $t_{\mathrm{start}}=1.5\tau_{\mathrm{B}}$. b) Density profiles
at some characteristic times for $K=1.4$. The washboard potential is indicated by the dotted lines.}
\label{control3}
\end{figure}

The control scheme considered so far focuses on the average particle position.
However, given that the basic dynamical variable in DDFT is the 
density profile $\rho(z,t)$, it is interesting to briefly discuss a control loop based on that quantity. Specifically, we consider the potential
\begin{equation}
\label{u_control3}
U^{\mathrm{c}}_2(z,[\rho])=-K_0z\left (1-\tanh\left[\rho(z,t-\tau)-\rho(z,t)\right]\right).
\end{equation}
The impact of this scheme on the function $\langle z\rangle_t$ is shown in fig.~\ref{control3}a, where the inset contains data for the weakly interacting
case $\varepsilon=1$.
For small control amplitudes ($K<1$), the behavior of both systems, $\varepsilon=4$ and $\varepsilon=1$, is similar to what is observed with the previous control loop (eq.~(\ref{f_control1})) in the sense that
the long-time velocity decreases with increasing $K$. However, contrary to this previous loop, further increase of $K$
then yields an {\em abrupt} reversal of the motion, that is, the particles
''bounce backwards'' in the sense that $v\rightarrow -\infty$. Moreover, this abrupt change occurs
at values of $K$ much smaller than the biasing force ($F=3$). The spatio-temporal behavior of the microscopic density profile $\rho(z,t)$
related to the sudden reversal of motion is illustrated in fig.~\ref{control3}b. At $t=3.5\tau_{\mathrm{B}}$, that is, before the reversal, the initial density peak centered
at $z=0$ has extensions to the next potential wells at positive $z$, indicating that the particles move rightwards. However, already at this time, there is a small
bump at positions $z<0$ not present within the first control scheme (see fig.~\ref{control1}c). 
At time $t=5\tau_{\mathrm{B}}$ after the reversal this additional bump has increased significantly
along with a left-ward shift of the center of mass. 
In a future publication we will analyze in more detail to which extent
the sudden reversal and the associated complex behavior of $\rho(z,t)$ is a true instability. Nevertheless, already the results in fig.~\ref{control3} indicate that a local control scheme could be more efficient, in the sense that much smaller perturbations $K$ are required to yield current reversal, compared to control focussing on a space-averaged quantity.
\section{Concluding remarks}
In conclusion, we have demonstrated that the transport in an interacting, driven colloidal system can be efficiently manipulated
by time-delayed feedback control. Our control goal in the present context was to stop or reverse the motion
in a tilted washboard potential below the critical driving strength $F_{\mathrm{crit}}$ beyond which the washboard becomes ineffective. We have shown that this goal
can be achieved with different control schemes that involve
the same delay time $\tau$ (chosen equal to the intrinsic, Brownian time scale) but different control targets (average particle position $\langle z\rangle_t$ versus density profile 
$\rho(z,t)$). The fact that the schemes produce comparable results indicates a certain 
robustness of the feedback control method for transport phenomena in 1D systems,  consistent with earlier theoretical findings for ratchet systems 
\cite{cao04,craig08,lopez08,borromeo06} and for non-interacting particles
in tilted washboards \cite{hennig1,hennig2}. In that sense, our study also supports the more general perspective that time-delayed feedback control
can be extremely useful for the manipulation of non-linear systems \cite{Schoellbuch}.

An experimental realization of 
the present results seems possible with micron-sized colloidal particles in 1D tilted washboard potentials created by optical (laser) fields
\cite{blickle07,evsti08,lopez08}. For instance, 
the study \cite{evsti08} employs particles of size $\sigma=1.5\mu m$ in potentials with periodicity $\lambda=3.14\mu m\approx 2.1\sigma$, and similar dimensions
occur in \cite{lopez08}. The barrier heights in \cite{evsti08} are 
$U=4.5-11.5$, and the biasing forces are in the range $0\leq F\lessapprox 100$, suggesting that our parameters for the uncontrolled system ($\lambda=8\sigma$, $U=15$, $F=3$) are not unrealistic. With respect to control, we stress that in colloidal systems
the trajectories of the {\em individual} particles can be monitored with a video camera \cite{evsti08,lopez08} yielding, in principle, {\em both} the average particle position and the full density field as possible control target. Indeed, the average colloid position is also 
targeted in a recent experimental realization of a flashing ratchet with time-delayed feedback control \cite{lopez08}. Moreover, in these experiments
the delay of $\tau=5$ms (arising from the finite time
required to locate the colloids), is much smaller than the time scale for diffusion ($\approx 300$ms), consistent with our assumption
$\tau\leq \tau_{\mathrm{B}}$.

Beyond the actual physical behavior, our study also shows that delayed feedback control
can be conveniently implemented within the framework of DDFT, a recently established method to describe colloidal dynamics
based on the microscopic interactions, yet with close relations to mesoscopic models of continuous phases \cite{espanol09}.
As a side-result of our analysis, we note that the DDFT appears to correctly describe also the non-controlled case,
which has already been intensively investigated by other methods \cite{marche09,risken}.
Since DDFT is an approximate theory, the present results for the impact of control 
remain to be tested against quasi-exact data from 
Brownian dynamics computer simulations. However, given the good performance of the DDFT in other contexts \cite{rex08,royall07}
we expect our findings to be at least qualitatively right.

Clearly, there is a number of questions prompted by our study. First, one needs to explore more systematically the precise role of the time delay in the
various control schemes, and of its interplay with the intrinsic timescales of the system. Second, given the complex behavior of non-controlled systems
at driving strength $F\approx F_{\mathrm{crit}}$ (related to ''giant diffusion") \cite{reimann01,evsti09,costantini99}, it would be interesting to extend the present analysis (where $F< F_{\mathrm{crit}}$) accordingly. Third, we need to better understand the role of the {\em nature} of the interactions
in the context of control. Here we have focused on soft, repulsive interactions characterizing, e.g., polymer coils \cite{likos01}, but also paramagnetic particles oriented parallel by a magnetic field (in 1D). What would happen in the presence of additional attractive contributions such as van-der-Waals- or depletion interactions (arising from solvent particles in the colloidal suspension)? Indeed, as
recently shown for colloids in 1D ratchet potentials (without control) \cite{pototsky10}, attractive interactions and the resulting
clustering tendency
strongly influences the transport behavior. Moreover, on a macroscopic scale, attractive interactions determine the {\em wettability} and as a consequence, the transport 
behavior of drops on (heterogeneous) substrates; a topic which is immediately relevant in the context of microfluidics \cite{john07} and may be accessible
within the present framework after a proper re-interpretation of the density fields.
Finally, the present methodology may also be used
to investigate systems consisting of several species, a question
relevant in the context of particle sorting effects \cite{ros05,eichhorn10}.
Work in these directions is under way.


\end{document}